

\frenchspacing

\parindent15pt

\abovedisplayskip4pt plus2pt
\belowdisplayskip4pt plus2pt
\abovedisplayshortskip2pt plus2pt
\belowdisplayshortskip2pt plus2pt

\font\twbf=cmbx10 at12pt
 at12pt
 at12pt

\font\sc=cmcsc10

\font\ninerm=cmr9
\font\nineit=cmti9
\font\ninesy=cmsy9
\font\ninei=cmmi9
\font\ninebf=cmbx9

\font\sevenrm=cmr7

\font\seveni=cmmi7
\font\sevensy=cmsy7

\font\fivenrm=cmr5
\font\fiveni=cmmi5
\font\fivensy=cmsy5

\def\nine{%
\textfont0=\ninerm \scriptfont0=\sevenrm \scriptscriptfont0=\fivenrm
\textfont1=\ninei \scriptfont1=\seveni \scriptscriptfont1=\fiveni
\textfont2=\ninesy \scriptfont2=\sevensy \scriptscriptfont2=\fivensy
\textfont3=\tenex \scriptfont3=\tenex \scriptscriptfont3=\tenex
\def\rm{\fam0\ninerm}%
\textfont\itfam=\nineit
\def\it{\fam\itfam\nineit}%
\textfont\bffam=\ninebf
\def\bf{\fam\bffam\ninebf}%
\normalbaselineskip=11pt
\setbox\strutbox=\hbox{\vrule height8pt depth3pt width0pt}%
\normalbaselines\rm}

\hsize30cc
\vsize44cc
\nopagenumbers

\def\luz#1{\luzno#1?}
\def\luzno#1{\ifx#1?\let\next=\relax\yyy
\else \let\next=\luzno#1\xxx\fi\next}
\def\sp#1{\def\xxx{\kern1.7pt}\def\yyy{\kern-1.7pt}\luz{#1}}
\def\spa#1{\def\xxx{\kern1pt}\def\yyy{\kern-1pt}\luz{#1}}

\newcount\beg
\newbox\aabox
\newbox\atbox
\newbox\fpbox
\def\abbrevauthors#1{\setbox\aabox=\hbox{\sevenrm\uppercase{#1}}}
\def\abbrevtitle#1{\setbox\atbox=\hbox{\sevenrm\uppercase{#1}}}
\long\def\pag{\beg=\pageno
\def\leftheadline{\noindent\rlap{\nine\folio}\hfil\copy\aabox\hfil}
\def\rightheadline{\noindent\hfill\copy\atbox\hfill\llap{\nine\folio}}
\def\phead{\setbox\fpbox=\hbox{\sevenrm
************************************************}%
\noindent\vbox{\sevenrm\baselineskip9pt\hsize\wd\fpbox%
\centerline{***********************************************}

\centerline{BANACH CENTER PUBLICATIONS, VOLUME **}

\centerline{INSTITUTE OF MATHEMATICS}

\centerline{POLISH ACADEMY OF SCIENCES}

\centerline{WARSZAWA 19**}}\hfill}
\footline{\ifnum\beg=\pageno \hfill\nine[\folio]\hfill\fi}
\headline{\ifnum\beg=\pageno\phead
\else
\ifodd\pageno\rightheadline \else \leftheadline \fi
\fi}}

\newbox\tbox
\newbox\aubox
\newbox\adbox
\newbox\mathbox

\def\title#1{\setbox\tbox=\hbox{\let\\=\cr
\baselineskip14pt\vbox{\twbf\tabskip 0pt plus15cc
\halign to\hsize{\hfil\ignorespaces \uppercase{##}\hfil\cr#1\cr}}}}

\newbox\abbox
\setbox\abbox=\vbox{\vglue18pt}

\def\author#1{\setbox\aubox=\hbox{\let\\=\cr
\nine\baselineskip12pt\vbox{\tabskip 0pt plus15cc
\halign to\hsize{\hfil\ignorespaces \uppercase{\spa{##}}\hfil\cr#1\cr}}}%
\global\setbox\abbox=\vbox{\unvbox\abbox\box\aubox\vskip8pt}}

\def\address#1{\setbox\adbox=\hbox{\let\\=\cr
\nine\baselineskip12pt\vbox{\it\tabskip 0pt plus15cc
\halign to\hsize{\hfil\ignorespaces {##}\hfil\cr#1\cr}}}%
\global\setbox\abbox=\vbox{\unvbox\abbox\box\adbox\vskip16pt}}

\def\mathclass#1{\setbox\mathbox=\hbox{\footnote{}{1991 {\it Mathematics
Subject
Classification}\/: #1}}}

\long\def\maketitlebcp{\pag\unhbox\mathbox
\footnote{}{The paper is in final form and no version
of it will be published elsewhere.}
\vglue7cc
\box\tbox
\box\abbox
\vskip8pt}

\long\def\abstract#1{{\nine{\bf Abstract.}
#1

}}

\def\section#1{\vskip-\lastskip\vskip12pt plus2pt minus2pt
{\bf #1}}

\long\def\th#1#2#3{\vskip-\lastskip\vskip4pt plus2pt
{\sc #1} #2\hskip-\lastskip\ {\it #3}\vskip-\lastskip\vskip4pt plus2pt}

\def\Proof{\vskip-\lastskip\vskip4pt plus2pt
\sp{Proo{f.}\ }\ignorespaces}

\def\endproof{\nobreak\kern5pt\nobreak\vrule height4pt width4pt depth0pt
\vskip4pt plus2pt}

\newbox\refbox
\newdimen\refwidth
\long\def\references#1#2{{\nine
\setbox\refbox=\hbox{\nine[#1]}\refwidth\wd\refbox\advance\refwidth by 12pt%
\def\textindent##1{\indent\llap{##1\hskip12pt}\ignorespaces}
\vskip24pt plus4pt minus4pt
\centerline{\bf References}
\vskip12pt plus2pt minus2pt
\parindent=\refwidth
#2

}}

\def\footnoterule{\kern -3pt \hrule width 4cc \kern 2.6pt}

\catcode`@=11
\def\vfootnote#1%
{\insert\footins\bgroup\nine\interlinepenalty\interfootnotelinepenalty%
\splittopskip\ht\strutbox\splitmaxdepth\dp\strutbox\floatingpenalty\@MM%
\leftskip\z@skip\rightskip\z@skip\spaceskip\z@skip\xspaceskip\z@skip%
\textindent{#1}\footstrut\futurelet\next\fo@t}
\catcode`@=12


\def\Q{{\cal Q}}
\def\A{{\cal A}}
\def\G{{\cal G}}
\def\D{{\cal D}}
\def\R{{\cal R}}
\def\H{{\cal H}}
\def\b{\bar}
\def\Sym{{\cal S}}
\def\reals{{\bf R}}             
\def\integers{{\bf Z}}          
\def\Sb{\bar{\Sigma}}
\def\lr{\leftrightarrow}
\font\bigmath=cmsy10 scaled \magstep 4
\def\bigast{\mathop{\vphantom{\sum}%
                    \lower2.5pt\hbox{\bigmath\char3}}}

\def\mapright#1{\smash{\mathop{\longrightarrow}\limits^{#1}}}
\def\maprights#1{\smash{\mathop{\rightarrow}\limits^{#1}}}
\def\maprightss#1{\mathop{\!\!\rightarrow\!\!}\limits^{#1}}
\def\mapdown#1{\Big\downarrow
  \rlap{$\vcenter{\hbox{$\scriptstyle#1$}}$}}
\def\diag{
  \def\normalbaselines{\baselineskip2pt \lineskip3pt
    \lineskiplimit3pt}
  \matrix}


\mathclass{Primary 83C99; Secondary 57R50, 55S37}
\abbrevauthors{D. Giulini}
\abbrevtitle{Group of Large Diffeomorphisms}

\title{The Group of Large Diffeomorphisms\\
       in General Relativity}

\author{D. Giulini}
\address{Fakult\"at f\"ur Physik der Universit\"at Freiburg,   \\
        Hermann Herder Stra\ss e 3, D-79104 Freiburg, Germany \\
   \\
        E-mail: giulini@sun2.ruf.uni-freiburg.de}

\maketitlebcp

\footnote{}{This research was supported in parts by the Isaac Newton
Institute, Cambridge (England), and the Center of Geometry and
Physics at the Pennsylvania State University. I thank these
institutions and the Banach Center in Warsaw for their
hospitality.}

\abstract{
We investigate the group of large diffeomorphisms fixing a
frame at a point for general closed 3-manifolds. We derive
some general structural properties of these groups which
relate to the picture of the manifold as being composed of
extended `objects' (geons).
}

\section{1. Introduction.}
The dynamics of General Relativity can be cast into the form of a
constraint Hamiltonian system. For this, the space-time $M$ is
assumed to be a topological product $\Sigma\times \reals$.
Restricting to pure gravity, the unconstrained phase space is
given by the cotangent bundle,
$T^*(\R)$, over the space $\R=\hbox{Riem}(\Sigma)$ of Riemannian
metrics on a 3-manifold $\Sigma$. Initial data thus consist of
a 3-metric, $g_{ab}$, and a contravariant tensor density of
weight one, $\pi^{ab}$. Once the spacetime $M$ has evolved from
$\Sigma$, $\pi^{ab}$ is related to the extrinsic curvature $K^{ab}$
of $\Sigma$ in $M$ via
$\pi^{ab}=\sqrt{g}(K^{ab}-g^{ab}K^c_c)$, where $g=\hbox{det}\{g_{ab}\}$.
But the data cannot be prescribed arbitrarily: they have to
satisfy the two constraints
$$\eqalignno{
 & D^a:=-2\nabla_b\pi^{ab} = 0\,,                   & (1.1)\cr
 & H  :=G_{abcd}\pi^{ab}\pi^{cd}-\sqrt{g} R    \,,  & (1.2)\cr}
$$
where $\nabla$ denotes the Levi-Civita connection for $g_{ab}$ and
$R$ its Ricci-scalar. $G_{abcd}$ is a symmetric, non-degenerate
bilinear form of signature type $(-,+,+,+,+,+)$, called the
DeWitt metric. Explicitly it reads:
$$
G_{abcd}={1\over 2\sqrt{g}}
\left(g_{ac}g_{bd}+g_{ad}g_{bc}-g_{ab}g_{cd}\right)\,.
\eqno{(1.3)}
$$
Upon Integration over $\Sigma$ it defines a (weakly) non-degenerate
form on $T^*(\R)$. Its metric properties on the subspace of
cotangent vectors satisfying $(1.1)$ are quite complicated.
In particular, it is neither positive definite nor even of definite
signature [1]. To set up initial data we now have to
1.) pick a 3-manifold $\Sigma$ and  2.) find a pair
$(g_{ab},\pi^{ab})$ satisfying $(1.1-2)$. It is known that
$(1.1-2)$ have solutions on all topologies $\Sigma$ [2].
In principle we therefore have to consider all 3-manifolds.

This setting is similar to the Hamiltonian formulation of
Yang-Mills theories: The space of gauge potentials $\A$
there corresponds to the space of metrics $\R$  here,
and the group of gauge transformations (vertical bundle
automorphisms) $\G$ there corresponds to the
diffeomorphism group $\D$ here. The constraint $(1.1)$,
called the diffeomorphisms constraint, is analogous to
the Gau\ss\ constraint in Yang-Mills theories. It should
be read as a momentum mapping for the action of the
diffeomorphism group on phase space which can be thought
of as the lift of the action on configuration space:
$$
\D\times\R\rightarrow \R\,,\quad (\phi,g_{ab})
          \mapsto \phi^* g_{ab}\,.
\eqno{(1.4)}
$$
A curve $t\rightarrow \phi_t$ in $\D$, where $\phi_0=id$,
defines a vector field $\xi$ on $\Sigma$ by
$\xi(p)={d\over dt}\vert_{t=0}\phi(p)$. The infinitesimal
version of $(1.4)$ is then given by
$$
  {d\over dt}\Big\vert_{t=0}\phi^*_t g_{ab}
  = \nabla_a\xi_b+\nabla_b\xi_a \,,
\eqno{(1.5)}
$$
where the right hand side is just the Lie derivative expressed in
terms of $\nabla$. The momentum map, $P$, applied to $\xi$, is
simply obtained by contracting the canonical 1- form (``$pdq$'')
with the right hand side of $(1.5)$:
$$
P(\xi)=2\int_{\Sigma}d^3x\, \pi^{ab}\nabla_b\xi_a
      = \int_{\Sigma}d^3x\,D^a\xi_a
      + 2\int_{\partial\Sigma}d^2x\,\xi_a\pi^{ab}n_b \,,
\eqno{(1.6)}
$$
where $d^3x$ and $d^2x$ denote coordinate volume elements and $n^a$
the outward pointing normal of $\partial\Sigma$. We thus see that
the diffeomorphism constraint generates those diffeomorphisms for
which the surface integral in $(1.6)$ vanishes. For example,
if $\Sigma$ has only a single connected boundary in an asymptotically
flat region and if $\xi$ asymptotically tends to a translation or
rotation, then the surface integral just corresponds to the total
momentum or angular momentum respectively. The corresponding motions
are {\it not} generated by the constraints. Only the asymptotically
trivial ones are.

{} From now on we restrict $\Sigma$ to have one regular end, i.e.,
there exists a compact set $K\subset\Sigma$ whose complement is
homeomorphic to the complement of a 3-ball in $\reals^3$.
This idealizes the case of an isolated gravitational system.
The regularity condition allows to 1-point compactify $\Sigma$ to
a compact manifold $\Sb=\Sigma\cup\infty$, where the added point
is called $\infty$. Although physically we are interested in
$\Sigma$ it is easier, and for our purposes admissible, to carry out
all constructions on $\Sb$. We thus take $\R$ to be the space of
3-metrics on $\Sb$, and $\D(\Sb)$ its diffeomorphism group.
Unless stated otherwise, we understand this group to consist of
{\it orientation preserving} diffeomorphisms only.
$\R(\Sb)$ is topologically a positive open cone in a vector space and
hence contractible. Since the diffeomorphism group `looks' differently
for different $\Sb$ it is given the argument to indicate
this dependence. There are two more diffeomorphism groups that we
will be interested in:
$$\eqalignno{
  \D_{\infty}(\Sb) &= \{\phi\in\D(\Sb)\ /\ \phi(\infty)=\infty\}\,,
&(1.7)\cr
  \D_F(\Sb) &= \{\phi\in\D_{\infty}(\Sb)\ /\ \phi_*\vert_{\infty}
                  =\hbox{id}\}\,.
&(1.8)\cr}
$$

Topologically we do not loose anything if we replace $\D(\Sigma)$
with $\D_{\infty}(\Sb)$ and the asymptotically trivial
diffeomorphisms of $\Sigma$ with $\D_F(\Sb)$. These spaces are
pairwise homotopy equivalent. The diffeomorphism
constraint generates the identity component of asymptotically trivial
diffeomorphisms, i.e. the identity component $\D^0_F(\Sb)$ of
$\D_F(\Sb)$. (Generally we indicate identity components by an upper
case 0.). The Hamiltonian formalism requires to regard
two points in the same orbit of the group generated by the
diffeomorphism constraint as physically identical. But there is still
an action of the quotient group,
$$
\Sym(\Sb)=\D_F(\Sb)/\D^0_F(\Sb)\,.
\eqno{(1.9)}
$$
Strictly speaking it is not required by the Hamiltonian formalism
to also identify $\Sym(\Sb)$-equivalent points. This is completely
analogous to the situation in Yang-Mills theories, where `large'
gauge transformations are not generated by the Gau\ss\ constraint.
This might lead to interesting physical consequences in the
quantum theory. An example is discussed in [3]. In the classical
theory it would be appropriate to quotient by the $\Sym(\Sb)$-action
if we agreed that the functions on phase space which we want to use
as observables cannot separate any two points in one orbit of
$\Sym(\Sb)$. This is in fact usually assumed, so that the classical
configuration space is then given by $\Q(\Sb)=\R(\Sb)/\D_F(\Sb)$,
which is in fact the base of a $\D_F(\Sb)$-principal-fibre-bundle [4]:
$$
\diag{\D_F(\Sb)&\mapright{}&\R(\Sb)\cr
      &&\mapdown{}\cr
      &&\Q(\Sb)\cr}
\eqno{(1.10)}
$$
The contractibility of $\R(\Sb)$ in the associated exact
sequence for homotopy groups directly leads to
$$\eqalignno{
&\pi_1\left(\Q(\Sb)\right)\cong\D_F(\Sb)/\D_F^0(\Sb) \,,
&(1.11)\cr
&\pi_n\left(\Q(\Sb)\right)\cong\pi_{n-1}\left(\D_F(\Sb)\right)
  =\Sym(\Sb) \,,\quad\forall n>1\,.
&(1.12)\cr}
$$
Some calculations and results of higher homotopy groups (1.12)
may be found in [5]. In this paper we are
concerned with the groups in (1.11).

If we did not quotient by $\Sym(\Sb)$ we would have obtained as
true configuration space the universal cover $\b {\Q}(\Sb)$ of
$\Q(\Sb)$. A classically equivalent procedure is to work with
$\b {\Q}(\Sb)$ and adding the requirement that all observables
must commute with the action of $\Sym(\Sb)$. This formulation
allows us to stick with the simply connected configuration space
$\b {\Q}(\Sb)$. In finite dimensions we would transfer this to
the quantum theory as follows [6]: take as Hilbert space $\H$
the $L^2$-space on $\b{\Q}(\Sb)$ with some $\Sym(\Sb)$-invariant
measure. Take as algebra of observables the commutant  of
$\Sym(\Sb)$. This induces a superselection structure with sectors
classified by the inequivalent irreducible unitary representations
of $\Sym(\Sb)$. For this reasons, and for the obvious reasons in the
classical theory, we are interested in the group $\Sym(\Sb)$ and
its representations.

Obviously, the quantum mechanical motivation as outlined above
has at least two serious problems. The first is that the
infinite dimensional space $\b{\Q}(\Sb)$ is almost certainly not
an appropriate measure space and therefore cannot be used to construct
$\H$. It would have to be replaced by something else which densely
contains $\b{\Q}(\Sb)$ as proper subset. This goal is pursued in
the works of Ashtekar and collaborators [7].
But what persists to happen is that $\Sym(\Sb)$ acts on the Hilbert
space and is required to commute with all observables. Our motivation
is therefore still valid. The second difficulty is that the discrete
group $\Sym(\Sb)$ is generically infinite and non-abelian so that
we cannot count on a general representation theory. Note that this is
a major difference to the situations usually encountered in Yang-Mills
theories. There the group of large gauge transformations is typically
abelian and the representation theory is simple. For example, the
famous $\theta$-angle just labels the irreducible unitary
representations of $\integers$. Some equivalents of the
$\theta$-angle in quantum-gravity have been studied in the
literature [8][9]. In the sequel of this paper we want to report
on some approaches and results to understand the general structure
of the group $\Sym(\Sb)$.

\section{2. 3-manifolds.}
We consider a general closed, connected, and oriented 3-manifold
$\Sb$. Up to permutation of factors it can be uniquely written as
a finite connected sum of prime 3-manifolds, simply called primes
(see [10] for more information on the general material that
follows; we denote the connected sum by $\uplus$):
$$
\Sb = \biguplus_{i=1}^{n+l} \Pi_i
    = \left(\biguplus_{i=1}^n P_i\right)    \uplus
    \left(\biguplus_{i=1}^l S^2\times S^1\right)\,,
\eqno{(2.1)}
$$
where amongst the general primes, $\Pi_i$, we have notationally
separated the irreducible ones, $P_i$, from the `handles',
$S^2\times S^1$. Recall that a 3-manifold is called irreducible
if every embedded 2-sphere bounds a 3-ball. Irreducibility implies
a trivial second homotopy group. The handle is the only reducible
prime and connected sums are of course always reducible. The known
primes with finite fundamental group are all of the form $S^3/G$
where $G$ is a finite subgroup of $SO(4)$ with proper action on
$S^3$. The irreducible ones with infinite fundamental group are
$K(\pi,1)$ spaces, i.e., their higher-than-first homotopy groups
are all trivial. Amongst those is the much better studied
subclass of sufficiently large $K(\pi,1)$'s, i.e. those which
contain incompressible surfaces. Simple examples are those of
the form $S^1\times R_g$, where $R_g$ is a genus $g$ surface, or
the six orientable flat 3-manifolds. It is conjectured but unproven
that all primes with finite fundamental group are covered by $S^3$
and all the irreducible ones with infinite fundamental group by $R^3$.
In (2.1) each $P_i$ is connected to the rest of $\Sb$ along a
2-sphere, $S_i$, $1\leq i\leq n$, whereas each handle is connected
along two 2-spheres, $S_{ij}$, $1\leq i\leq l$, $j=1,2$, which
should be thought of as the boundaries of the cylinder
$[0,1]\times S^2$ [11][12].

The fundamental group of the connected sum is
given by the free product of the individual ones:
$$
\pi_1(\Sb,\infty)=\left(\bigast_{i=1}^n\pi_1(P_i)\right)*
                \left(\bigast_{i=1}^l\integers\right)\,.
\eqno{(2.2)}
$$

It is useful to visualize $\Sb$ in terms of the connected sum (2.1).
Roughly speaking, $\Sb$ is divided into the regions interior to the
primes (called interior regions) and their complement (called exterior
region). The interior regions are connected to the exterior region
along the pairwise disjoint 2-spheres $S_i$ and $S_{ij}$.
This is explained in more detail in [11][12].
This allows to divide the diffeomorphisms into three different
types: 1.) those with support in the interior
region (called interior diffeomorphisms), 2.) those which leave the exterior
and interior region setwise invariant (called exterior
diffeomorphisms), like the exchange of two diffeomorphic primes,
and 3.) special diffeomorphisms not in 1.) and 2.), called exterior
slides. These will be explained below.

An important tool in analyzing $\Sym(\Sb)$ is the following
homomorphism
$$
h_{\infty}(\Sb):\D_{\infty}(\Sb)/\D_{\infty}^0(\Sb)
           \rightarrow \hbox{Aut}(\pi_1(\Sb,\infty))\,,
\eqno{(2.3)}
$$
which one obtains by composing the loop $\gamma$ that represents
the element in the homotopy group with $\phi\in\D_{\infty}(\Sb)$
to $\phi\circ\gamma$. (This also works for higher homotopy groups.)
Useful information on $\Sym(\Sb)$ is then obtained by investigating
the kernel and image of $h_{\infty}(\Sb)$. For later notational
convenience we also introduce the map $h_F$ which is defined just
like $h_{\infty}$ but with $\D_F/\D_F^0$ replacing
$\D_{\infty}/\D_{\infty}^0$.

Let us describe some of the important diffeomorphisms.
First we turn to the internal diffeomorphisms. Take an inner collar
neighbourhood of $S_i$, and define a diffeomorphisms by a
relative $2\pi$-rotation of its boundary 2-spheres, and extend
it to the rest of the manifold by the identity. This is called a
{\it rotation of} $P_i$. If this diffeomorphism is not isotopic
to the identity within the class of inner diffeomorphisms of
$P_i$ holding $S_i$ fixed, we call $P_i$ {\it spinorial}.
Amongst the known primes all but the lens spaces and the handle
are spinorial and a connected sum is spinorial iff at least one
prime is [12]. For spinorial manifolds $\Sym(\Sb)$ is a central
$\integers_2$-extension of $\D_{\infty}(\Sb)/\D_{\infty}^0(\Sb)$.
In the non-spinorial case they are isomorphic~[5]. The same
relation hold for the kernels of $h_F(\Sb)$ and $h_{\infty}(\Sb)$.
Similarly, for a handle we can define an internal
diffeomorphism by a relative $2\pi$-rotation of $S_{i1}$ and
$S_{i2}$, extended by the identity. This is called a {\it twist}
of the $i$'th handle. It is not isotopic to the identity and
generates $\Sym(S^2\times S^1)$.
Rotations of spinorial primes and twists each generate a
$\integers_2$ subgroup of $\Sym(\Sb)$. Moreover, since they
rotate parallel to 2-spheres, they are in the kernel of
$h_{\infty}(\Sb)$. Clearly
any two internal diffeomorphisms of different primes commute.
Hence there is a  homomorphism of the direct product of all
$\Sym(\Pi_i)$ into $\Sym(\Sb)$. Hendriks and McCullough
have proven that this homomorphism is in fact an injection [13]:
$$
I:\,\prod_{i=1}^n \Sym(P_i)\times \integers_2^l
                  \hookrightarrow\Sym(\Sb)\,.
\eqno{(2.4)}
$$
Applying a more general result of McCullough's [14]
to the present situation shows that the kernel of $h_{F}(\Sb)$
lies in the image of the map $I$, that is
$$
\hbox{ker}((h_F(\Sb))=\left(\prod_{i=1}^{l} \integers_2\right)
           \times \prod_{i=1}^nK_i\,,
\eqno{(2.5)}
$$
where $K_i$ denotes the kernel of $h_F$ in $\Sym(P_i)$, which,
we recall, is a $\integers_2$-extensions of the kernel of
$h_{\infty}(P_i)$ in $\D_{\infty}(P_i)/\D_{\infty}^0(P_i)$ if $P_i$ is
spinorial, and isomorphic otherwise. On the other hand, suppose two
diffeomorphisms $d,d'\in\D_{\infty}(P_i)$ have the same
image under $h_{\infty}(P_i)$. Irreducible primes have $\pi_2=0$
and those of the form $S^3/G$ have $\pi_3=\integers$ whereas
$K(\pi,1)$ primes have $\pi_3=0$. This implies that $d$ and
$d'$ are homotopic (see e.g. Lemma 5.1 in [15]).
But for $P_i$ for which homotopy implies isotopy -- we call it the
HI-property and manifolds with this property HI-manifolds -- this
implies the triviality of $K_i$. No non-HI-prime is known
so far (though non-HI connected sums! [16]).
But some cases are undecided and we refer to theorem A1 in
[5] for a list of which primes have so far been
proven to be of HI type. For them we have the

\th{Proposition}{1.}{Let $\Sb$ be composed of HI-primes of which
$k$ are spinorial. Then
$\hbox{ker}(h_F(\Sb))= \prod_{i=1}^{k+l} \integers_2$, generated
by rotations of spinorial primes and twists of handles.}

The external diffeomorphism which we have to mention are the
{\it exchanges} of diffeomorphic primes and the {\it spins}
of handles. The latter just correspond to a mutual exchange
of the two handle-ends. More details are given in [12].

Next we describe slide diffeomorphisms:
Given a loop $\gamma:[0,1]\ni s\rightarrow\Sb$ with
toroidal tubular neighbourhood $T\subset\Sb$ coordinatized by
$(r,\theta,\varphi)\in [0,2]\times [0,2\pi)\times[0,2\pi)$, where
$r$ is a radial coordinate parameterizing coaxial 2-tori $T_t=\{r=t\}$
with axis $T_0=\gamma$. $\theta$ and $\varphi$ coordinatize the
toroidal latitudes and longitudes respectively.
Let $\lambda:\reals\rightarrow [0,1]$ be a $C^{\infty}$ step function:
$\lambda(x)=0$ for $x\leq 0$, $\lambda(x)=1$ for $x\geq 1$. We define
a 1-parameter family of diffeomorphisms $\phi_t$ of $\Sb$ as follows:
$$
\phi_t(p)=
\cases{p
       & for $p\in \Sb-T\,,$\cr
       (r,\theta, \varphi+t2\pi \lambda(2-r))
       & for $p=(r,\theta,\varphi)\in T$.\cr}
\eqno{(2.6)}
$$
The diffeomorphism $\phi_1=\phi_{t=1}$ is the identity inside $T_1$.
It is called a {\it slide along} $\gamma$.
Suppose $\infty$ is on $\gamma$. Then $\phi_1\in \D_F(\Sb)$ and the
action of $\phi_1$ on $\pi_1(\Sb,\infty)$ is by conjugation with
$[\gamma]$, the class of $\gamma$ in $\pi_1(\Sb,\infty)$. This shows
that
$$
\hbox{Inn}(\pi_1(\Sb,\infty))\subset\hbox{Im}(h_{\infty}(\Sb))\,,
\eqno{(2.7)}
$$
where we could have replaced $h_{\infty}$ with $h_F$.
Next suppose we slightly change the construction
above in that one of the spheres $S_i$ (or $S_{ij}$) now sits inside
$T_1$. We then call $\phi_1$ a slide of $P_i$ (or the $j$'th end of
handle $i$) along $\gamma$. This diffeomorphism is isotopic to the
identity if $\gamma$ is contractible, but not otherwise [11].

An important theorem now states that $\Sym(\Sb)$ is generated by
internal, exchange, and slide diffeomorphisms. The slides we need to
consider are slides of primes or handle-ends along non-contractible
loops. Consequently, $\Sym(\Sb)$ is finitely generated if each
$\Sym(P_i)$ is. Lucid proofs of these important results may be
found in~[11].

The individual groups $\Sym(P_i)$ have been calculated for many
irreducible primes. See [15] for the spherical primes
$S^3/G$. In the next section we show how to obtain a general
result for the class of sufficiently large $K(\pi,1)$ primes.

\section{3. $\Sym(P)$ for $P$ sufficiently large $K(\pi,1)$.}
Let for the moment $P$ be an irreducible prime. The spaces
$\D_{\infty}(P)$ and $\D(P)$ are related by the fibration
$$
\diag{
\D_{\infty}(P)&\mapright{i}&\D(P)&&\cr
&&\mapdown{p}&&\quad p(\phi):=\phi(\infty)\cr
&& P &&\cr }
\eqno(3.1)
$$
whose associated exact sequence for homotopy groups ends with
(dropping the arguments $P$ for the moment)
$$
\diag{
\cdots & \maprightss{} & \pi_1(\D_{\infty}) & \maprights{}
       & \pi_1(\D)&\maprights{p_*} & \pi_1 & \maprights{\partial_*}
       & \pi_0(\D_{\infty}) & \maprights{i_*} & \pi_0(\D)
       & \maprights{} & 1\cr}
\eqno{(3.2)}
$$
The map $\partial_*$ is as follows: Take a class
$[\gamma]\in\pi_1(P,\infty)$ and a representing loop, $\gamma_t$, at
$\infty$. Take a lift, $\phi_t$, of $\gamma_t$ in $\D_{\infty}$
starting at the identity map, $\phi_0=id$, so that
$\phi_t(\infty)=\gamma_t$. The end point, $\phi_1$, defines a class
$[\phi_1]\in\pi_0(\D_{\infty}):=\D_{\infty}/\D_{\infty}^0$, and
one defines $\partial_*([\gamma]):=[\phi_1]$. This is indeed a
well defined map since it is independent of the representative
$\gamma_t$ of $[\gamma]$.

The lift $\phi_t$ may of course just be given by a slide along
$\gamma$, as defined in (2.3).
As noted above, $h_{\infty}([\phi_1])$ just corresponds to conjugation
by $[\gamma]$, so that $h_{\infty}\circ \partial_*=Ad$, where $Ad$
denotes the `adjoint-homomorphism' from $\pi_1$ to $\hbox{Inn}(\pi_1)$.
Its kernel is $C_{\pi}$, the centre of $\pi_1$. Exactness of (3.2)
implies that the image of $p_*$ lies within that kernel:
$$
\hbox{Im}p_*\subset C_{\pi}\,.
\eqno{(3.3)}
$$
This allows to write the inner automorphisms as a double quotient
$$
\hbox{Inn}(\pi_1)\cong {\pi_1\over \hbox{Im}p_*}\Big\slash
                 {C_\pi\over \hbox{Im}p_*}
                =:q\left(\pi_1/\hbox{Im}p_*\right)\,,
\eqno(3.4)
$$
where $q$ just denotes the quotient map with respect to
$C_\pi/\hbox{Im}p_*$. Let
$h:\D/\D^0\rightarrow\hbox{Out}(\pi_1)$ be
the non-based version of $h_{\infty}$. Diagram $(3.2)$ can now
be completed by the maps $q$, $h_{\infty}$ and $h$ with two
commutative squares:
$$
\diag{
1 & \mapright{} & {\pi_1\over {\rm Im}p_*} & \mapright{\partial_*}
  & \pi_0(\D_{\infty}) & \mapright{i_*} & \pi_0(\D)
  & \mapright{} & 1                                         \cr
 && \mapdown{q} && \mapdown{h_{\infty}} && \mapdown{h} &&   \cr
1 & \mapright{} & {\rm Inn}(\pi_1) & \mapright{i'} & {\rm Aut}(\pi_1)
  & \mapright{p'} & {\rm Out}(\pi_1) & \mapright{} & 1            \cr}
\eqno{(3.5)}
$$
Commutativity of the left square follows from the discussion above
whereas commutativity of the right square is obvious. $(3.3)$ implies
that $q$ is surjective which implies for the diagram that
$h_{\infty}$ is surjective if and only if $h$ is. If we now restrict
$P$ to be an irreducible HI-prime we can to show

\th{Lemma}{1.}{Let $P$ be an irreducible HI-prime, then $q$ is an
               isomorphism}

\Proof
For irreducible HI primes $h_{\infty}$ is injective.
{} From $(3.5)$ $h_{\infty}\circ\partial_*=i'\circ q$
with injective maps $h_{\infty}$, $\partial_*$ and
$i'$. So $q$ must be injective. But $q$ is a quotient
map with respect to $C_{\pi}/\hbox{Im}p_*$. So
$\hbox{Im}p_*=C_{\pi}$ and $q$ is an isomorphism$\bullet$
\vskip4pt plus 2pt

Hatcher has proven [17] that for sufficiently large
$K(\pi,1)$ primes (which all have the HI property)
$h$ is an isomorphism if $\D$ includes orientation reversing
diffeomorphisms (if existent). Generalizing for the moment
to this case, diagram $(A.5)$ then implies
that $h_{\infty}$ is also an isomorphism. Taken together
with the fact that $K(\pi,1)$ primes are spinorial (this
is e.g. shown in [18] or [19]) this implies

\th{Proposition}{2.}{Let $P$ be a sufficiently large $K(\pi,1)$ prime.
                   Then $\Sym(P)$ is a central $\integers_2$
                   extension of $\hbox{Aut}^+(\pi_1(P))$}.

Here we had to account for the possibility that $P$ allows for
orientation reversing diffeomorphisms all of which act
non-trivially on the fundamental group. Since diffeomorphisms
in $\D_F(P)$ are necessarily orientation preserving, we can in this
case only reach an index 2 subgroup of $\hbox{Aut}(\pi_1(P))$ which
we denoted by an upper case $+$ sign. This difference does not exist
if $P$ does not allow for orientation reversing diffeomorphisms
(this is the generic case, see e.g. [12] for collective
information on this point), or if there exists an orientation
reversing diffeomorphism acting trivially on the fundamental group,
like e.g. for the handle.
For example, for the 3-torus $T^3$ we have
$\hbox{Aut}(\integers^3)=GL(3,\integers)$,
$\hbox{Aut}^+(\integers^3)=SL(3,\integers)$ and one may show that
$\Sym(T^3)$ is given by the Steinberg group $St(3,\integers)$
(see [20] for more information on the Steinberg groups).

\section{4. General Remarks.}
For a physicist it is tempting to think of the manifold $\Sb$ as being
composed of elementary objects, the primes, just like a collection
of $N=n+l$ particles from $d$ different species each with its
own characteristic internal symmetry group $G_r$, $1\leq r\leq d$
(in doing this we think of the spin of a handle as internal
diffeomorphism). In this analogy diffeomorphic primes correspond to
particles of one species and the `internal' symmetry groups $G_r$ to
the groups $\Sym(\Pi_r)$. Let there be $n_r$ primes in the $r$'th
diffeomorphism class so that $\sum_{r=1}^d n_r =N$. In the particle
picture the total symmetry group would be a semidirect product
of the internal symmetry group $G^I$ with the external symmetry group
$G^E$. These are give by:
$$\eqalignno{
G^I = &\prod_{i=1}^{n_1}G_1\times\cdots\times \prod_{i=1}^{n_d}G_d\,,
      &(4.1)\cr
G^E = & S_{n_1}\times\cdots\times S_{n_d}\,.
      &(4.2)\cr}
$$
There is a homomorphism $\theta:G^E\rightarrow\hbox{Aut}(G^I)$, defined
by $\theta=\theta_1\times\cdots\times\theta_d$, where
$$\eqalign{
\theta_j:\,S_{n_j}
& \rightarrow \hbox{Aut} \left(\prod_{i=1}^{n_j}G_j\right)\,, \cr
  \sigma
& \mapsto \theta_j(\sigma):\,(g_1,\dots ,g_{n_j})\mapsto
             (g_{\sigma(1)},\dots ,g_{\sigma(n_j)})\,.  \cr}
\eqno{(4.3)}
$$
The semidirect product $G^I\times_{\theta}G^E=:G^P$, which we call
the {\it particle group}, is now defined via the multiplication law as
follows: Let $\gamma_j\in\prod_{i=1}^{n_j}G_j$,
$j=1,...,d$, then
$$\eqalign{
 & \left(\gamma'_1,\dots ,\gamma'_d\,;\,\sigma'_1,\dots ,\sigma'_d\right)
   \left(\gamma_1, \dots ,\gamma_d \,;\,\sigma_1, \dots ,\sigma_d\right)\cr
=& \left(\gamma'_1[\theta_1(\sigma'_1)(\gamma_1)],\cdots ,
    \gamma'_d[\theta_d(\sigma'_d)(\gamma_d)]\,;\,
    \sigma'_1\sigma_1,\dots ,\sigma'_d\sigma_d \right)\,.\cr}
\eqno{(4.4)}
$$

{} From the discussion in section 2 it is clear that the group $G^P$ is a
subgroup of $\Sym(\Sb)$ generated by internal and external
diffeomorphisms. However, we also had to consider slides which were
neither internal nor external and which are not compatible with the
particle picture just used. As regards our understanding of $\Sym(\Sb)$,
it seems that one of the most interesting questions is to ask for the
r\^ole of slides in building up $\Sym(\Sb)$. For example, we may ask
for the normal subgroup $N_S$ generated by slides and whether it
overlaps with the particle group $G^P$. If so, the quotient
$\Sym(\Sb)/N_S$ will only be a factor group of $G^P$. On the other hand,
it may also be that $\Sym(\Sb)$ is a semidirect product of $N_S$ with
$G^P$. In the sequel of this paper we attempt to answer this question.

\section{5. Examples.}
In this section we wish to develop some feeling for the question just
posed by presenting two examples of manifolds each consisting
of a $N>2$ fold connected sum of diffeomorphic primes.
In the first example the primes will be handles:
$$
\Sb=\biguplus_{i=1}^N S^2\times S^1\,,
\eqno{(5.1)}
$$
so that the fundamental group of $\Sb$ is just the free group on
$N$ generators $g_1,\cdots, g_N$:
$$
\pi_1(\Sb)=\bigast_{i=1}^N {\integers} = F_N\,.
\eqno{(5.2)}
$$
We represent each $g_i$ by a loop based at $\infty$ which enters the
$i$'th handle through $S_{i1}$ and leave it through $S_{i2}$
(call it the positive direction).
The group $\hbox{Aut}(F_N)$ can be generated by the four operations
$$\eqalignno{
P&:[g_1,g_2,g_3,\dots ,g_N]\mapsto [g_2,g_1,g_3,\dots ,g_N],     &(5.3a)\cr
Q&:[g_1,g_2,g_3,\dots ,g_N]\mapsto [g_2,g_3,g_4,\dots ,g_N,g_1], &(5.3b)\cr
O&:[g_1,g_2,g_3,\dots ,g_N]\mapsto [g_1^{-1},g_2,g_3,\dots ,g_N],&(5.3c)\cr
U&:[g_1,g_2,g_3,\dots ,g_N]\mapsto [g_1g_2,g_2,g_3,\dots ,g_N],  &(5.3d)\cr}
$$
whose interpretation is as follows: $P$ exchanges handle no 1
and handle no 2, $Q$ exchanges all $N$ handles in cyclic order, $O$
spins handle no 1, and $U$ slides the second end of the first handle
through the second handle in a negative direction. This means that
by inspection we have just proven that in the present case the map
$h_{\infty}(\Sb)$ in (2.3) is surjective. From (2.5) we know that its kernel
and hence that $\Sym(\Sb)$ is an extension of $\hbox{Aut}(F_N)$ by
$\prod_{i=1}^N\integers_2$. It is also easy to describe the extension
explicitly. First of all, $P$ and $Q$ alone generate the permutation
group $S_N$ of $N$ objects which is a sub- but no factor-group of
$\hbox{Aut}(F_N)$, and $P,Q,O$ generate a subgroup of order $2^NN!$
whose interpretation is that of a particle group with internal
$\integers_2$ symmetry (the spin). A presentation of the full group
is obtained by adding a generator $T$ which generates a twist on the
first handle. It commutes with $U$ and $O$, squares to the identity,
and has the same relations with $P,Q$ as $O$ has. Since we are given
an explicit presentation of $\hbox{Aut}(F_N)$ in terms of
$P,Q,O,U$~[21] we can now give an explicit presentation of
$\Sym(\Sb)$ in terms of $P,Q,O,T,U$. We use the symbol $V$ to
denote either $O$ or $T$ so that a relation containing $V$ is
meant to be valid with $V=O$ and $V=T$. $A\lr B$ says that $A$
commutes with $B$:
$$\eqalignno{
&P^2=V^2=E                                               &(5.4a)\cr
&(QP)^{N-1}=Q^N                                          &(5.4b)\cr
&P\lr Q^{-i}PQ^i,\quad\hbox{for}\ 2\leq i\leq N/2        &(5.4c)\cr
&V\lr Q^{-1}PQ                                           &(5.4d)\cr
&V\lr QP                                                 &(5.4e)\cr
&(PV)^4=E                                                &(5.4f)\cr
&T\lr O                                                  &(5.4g)\cr
&U\lr T                                                  &(5.4h)\cr
&U\lr Q^{-2}PQ^2\qquad\hbox{for}\ N>3                    &(5.4i)\cr
&U\lr QPQ^{-1}PQ                                         &(5.4j)\cr
&U\lr Q^{-2}OQ^2                                         &(5.4k)\cr
&U\lr Q^{-2}UQ^2\qquad\hbox{for}\ N>3                    &(5.4l)\cr
&U\lr OUO                                                &(5.4m)\cr
&U\lr PQ^{-1}OUOQP                                       &(5.4n)\cr
&U\lr PQ^{-1}PQPUPQ^{-1}PQP                              &(5.4o)\cr
&(PQ^{-1}UQ)^2=UQ^{-1}UQU^{-1}                           &(5.4p)\cr
&U^{-1}PUPOUOPO=E                                        &(5.4q)\cr}
$$

Most of the $\lr$ relations have a trivial meaning. For example,
(5.4j) just says that $U$ commutes with a cyclic permutation of
$g_i$'s for $i>2$. But this is obvious since their supports
may be chosen to be disjoint. The most remarkable relation
is (5.4q) which can be rewritten in the form
$OP=U^{-1}(PUP^{-1})(OUO^{-1})$. Hence $OP$ is in the normal
subgroup generated by slides. Adding the relation $U=E$ will
therefore not lead to the particle group. In fact we have

\th{Theorem}{1.}{The quotient with respect to the normal subgroup
generated by slides is $\integers_2\times \integers_2$ generated
by $T$ and $P$.}

\Proof Setting $U=E$ in (5.4q) leads via (5.4a) to $P=O$.
(5.4e) for $V=O$ then implies $P\lr Q$ and (5.4b) implies
$Q=P^{N-1}$, i.e., $Q=E$ for $N$ odd and $Q=P$ for $N$ even.
We are thus left with two involutive commuting generators $T$
and $P$~$\bullet$
\vskip4pt plus2pt

By simply checking the relations one may moreover prove [12]

\th{Theorem}{2.}{Given a representation $\rho$ of $\Sym(\Sb)$. The
following conditions are equivalent: 1.) $\rho$ is abelian,
2.) slides are represented trivially, 3.) $\rho$ correlates $P$ and
$O$, i.e. $\rho(P)=\rho(O)$, and 4.) slides and exchanges commute
under $\rho$.}

As regards this theorem we mention that the case $N=2$ behaves rather
differently~[12].

In our second example we take as prime the real projective space
$RP^3$. The setting is very similar to the previous case so that we
can be brief. The fundamental group is the free product on $N$
involutive generators $g_1,\cdots, g_N$, $g_j^2=E$. There is no
non-trivial internal diffeomorphism (e.g. [15]),
i.e. $\Sym(PR^3)=E$ so that the kernel of $h_{\infty}(\Sb)$
trivial. $\Sym(\Sb)$ is generated by the operations:
$$\eqalignno{
P&:[g_1,g_2,g_3,\dots ,g_N]\mapsto [g_2,g_1,g_3,\dots ,g_N],    &(5.5a)\cr
Q&:[g_1,g_2,g_3,\dots ,g_N]\mapsto [g_2,g_3,g_4,\dots ,g_N,g_1],&(5.5b)\cr
U&:[g_1,g_2,g_3,\dots ,g_N]\mapsto [g_2^{-1}g_1g_2,g_2,g_3,
                                              \dots ,g_N],      &(5.5c)\cr}
$$
and the relations are given by
$$
\eqalignno{
& P^2=U^2=E                                                       &(5.6a)\cr
& (QP)^{N-1}=Q^N=E                                                &(5.6b)\cr
& P\lr O^{-i}PQ^i,\quad\hbox{for}\, 2\leq i\leq N/2               &(5.6c)\cr
& U\lr Q^{-2}PQ^2,\quad\hbox{for}\, N>3                           &(5.6d)\cr
& U\lr QPQ^{-1}PQ                                                 &(5.6e)\cr
& U\lr Q^{-2}UQ^2                                                 &(5.6f)\cr
& U\lr Q^{-1}PUPQ                                                 &(5.6g)\cr
& Q^{-1}UQUQ^{-1} UQ=PQ^{-1}UQPUPQ^{-1}UQP,\quad\hbox{for}\,
                                                  N\geq 3\,.      &(5.6h)\cr}
$$
Of course, the first three relations again imply that $P,Q$ generate
the permutation group $S_N$ which here is identical to the particle
group $G^P$. All $\lr$ relations can be understood from
simple arguments saying that the diffeomorphisms representing the
left and right sides can be chose to have disjoint support. Relation
(5.6h) is less obvious. If $\mu_{ij}$ represents the unique slide of
prime $j$ through prime $i$, then it can be expressed in
the form: $\mu_{31}\mu_{21}\mu_{31}=\mu_{32}\mu_{21}\mu_{32}$.
(Relation (5.4p) has a similar interpretation using slides of
handle-ends.) In particular, it is
a relation purely amongst slides. Setting $U=E$ in (5.6h) leads
to no additional relations for $P$ and $Q$, so that now we indeed have
$\Sym(\Sb)/N_S=G^P$.

How can such a different behaviour be explained? Formally, a major
difference between handles and irreducible primes is that for handles
we have the possibility to slide only their ends rather than the
whole prime. This results is the difference of (5.3d) and (5.5c):
a conjugation on generators, like in (5.5c), can be generated from
(5.3) (sliding both ends) but a mere left or right multiplication on
generators , like in (5.3d), cannot be generated from (5.5).
This suggests that the presence of handles is responsible for the particle
group not being a quotient. In the next section we report some work
that proves this conjecture.

\section{6. The General Case.}
We recall that in the general case $\pi_1(\Sb)$ is a free product of
the groups $\pi_1(P_i)$ and some $\integers$'s. $\Sym(\Sb)$ is to
be built from $\hbox{ker}(h_{\infty}(\Sb))$ and
$\hbox{Im}(h_{\infty}(\Sb))\subseteq\hbox{Aut}(\pi_1(\Sb))$.
The reason why we can make statements about the general case lies in
the happy fact that we are explicitly given a presentation for the
automorphism group of free products. We use the form given in
[22] (called the Fuks-Rabinovitch presentation),
in which the generators of $\hbox{Aut}(\pi_1(\Sb))$ are given by
those of $\hbox{Aut}(\pi_1(P_i))$ and others which correspond to spin-,
exchange-, and slide-diffeomorphisms. The lack of surjectivity of
$h_{\infty}(\Sb)$ on $\hbox{Aut}(\pi_1(\Sb))$ resides entirely
in the lack of surjectivity of its restrictions $h_{\infty}(\Pi_i):
\Sym(\Pi_i)\rightarrow\hbox{Aut}(\pi_1(\Pi_i))$. For example,
using proposition 2, we may infer surjectivity if $\Sb$ is a
connected sum of handles and sufficiently large $K(\pi, 1)$ primes.
Since these are all HI-primes, we can combine this with the first
example of the previous section and infer

\th{Theorem}{3.}{Let $\Sb$ be a connected sum of $l$ handles and
$n$ sufficiently large $K(\pi,1)$ primes. Then $\Sym(\Sb)$ is an
extension of $\hbox{Aut}^+(\pi_1(\Sb))$ by
$\prod_{i=1}^{n+l}{\integers}_2$.}

Like in the $N$-handle example the extension would not be difficult
to describe. The $\integers_2$'s correspond to twists and rotations
which commute with other internal diffeomorphisms and slides, but
are acted upon by exchanges in the obvious way. Together with the
explicit presentation of $\hbox{Aut}(\pi_1(\Sb))$ this is then
sufficient information to write down a presentation of $\Sym(\Sb)$.

We now address the question posed at the end of the previous
section. To do this one has to explicitly go through the
Fuks-Rabinovitch relations as given in [22]. We will not repeat
them here but rather refer the reader to the article by
McCullough and Miller. We shall adopt their numbering of the
relations. Amongst these relations are four of particular interest
for us whose implications we now wish to explain:

{\it Relation 30}: Given any connected sum containing (amongst
possible other primes) two handles $H_i,H_j$ and another prime
$\Pi_k$ (which may be a handle or not), then slides of a handle-end
of $H_j$ through $\Pi_k$ can be written as a commutator
of slides.

{\it Relation 47}: Given any connected sum containing (amongst
possible other primes) a prime $\Pi_i$ (which may be a handle
or not), a handle $H_j$, and an irreducible prime $P_k$, then any
slide of $P_k$ through $\Pi_i$ can be written as a commutator
of slides.

These two observations imply

\th{Theorem}{4.}{If $\Sb$ contains at least three handles
then slides generate a perfect subgroup $N_S$ of $\Sym(\Sb)$.}

This generalizes our observation concerning slides made on
the first example above.

{\it Relation 31}: Given any connected sum containing (amongst
possible other primes) two handles $H_i$ and $H_j$, then a spin
of $H_j$ followed by an exchange of $H_i$ and $H_j$ is generated
by slides of handle-ends of $H_i$ through $H_j$ and of $H_j$
through $H_i$.

{\it Relation 48}: Given any connected sum containing (amongst
possible other primes) an irreducible prime $P_i$ and a handle
$H_j$, then internal slides of $P_i$ are generated by external
slides of $P_i$ through $H_j$ and ends of $H_j$ through $P_i$.

{\it General Observation}: The last two relations are the only
ones that imply additional relations for non-slide generators
when the slides generators are set equal to the identity $E$.
This implies:

\th{Theorem}{5.}{If $\Sb$ contains no handle in its prime
decomposition, then $\Sym(\Sb)$ divide by the normal subgroup
generated by slides $N_S$ is given by the `particle group' $G^P$.}

Note that this implies in particular that all (irreducible)
representations of $G^P$ extend to (irreducible) representations
of $\Sym(\Sb)$. We can now understand our two examples in the
previous section as extreme cases for the last two theorems.

\references{[22]}{{
\item{[1]}
D. \spa{Giulini},
{\it What is the geometry of superspace?}\/,
Phys.~Rev.~D~51 (1995), 5630--5635.

\item{[2]}
D. \spa{Witt},
{\it Vacuum space-times that admit no maximal slice}\/,
Phys.~Rev.~Lett.~57 (1986), 1386--1389.

\item{[3]}
D. \spa{Giulini},
{\it Asymptotic symmetry groups of long-ranged gauge configurations}\/,
Mod. Phys. Lett.~A~10 (1995), 2059--2070.

\item{[4]}
A.E. \spa{Fischer},
{\it Resolving the singularities in the space of Riemannian
geometries}\/,
Jour. Math. Phys.~27 (1986), 718--738.

\item{[5]}
D. \spa{Giulini},
{\it On the configuration space topology in general relativity}\/,
Helv.~Phys.~Acta~68 (1995), 87--111.

\item{[6]}
D. \spa{Giulini},
{\it Quantum mechanics on spaces with finite fundamental group}\/,
Preprint Freiburg THEP-95/16 and quant-ph 9508011.
Submitted for Publication.

\item{[7]}
A. \spa{Ashtekar}, J. \spa{Lewandowski}, D. \spa{Marolf}\/,
J. \spa{Mourao}, and T. \spa{Thiemann},
{\it Quantization of diffeomorphism invariant theories with local
degrees of freedom},
preprint UCSBTH-95-7 and gr-qc/9504018.

\item{[8]}
D. \spa{Giulini} and J. \spa{Louko},
{\it Theta-sectors in spatially  flat quantum cosmology}\/,
Phys.~Review~D~46 (1992), 4355--4364.

\item{[9]}
D. \spa{Giulini} and J. \spa{Louko},
{\it Diffeomorphism invariant states in Wittens 2+1 quantum
gravity on $R\times T^2$}\/,
Class.~Quant.~Grav. To appear.

\item{[10]}
J. \spa{Hempel},
{\it 3-Manifolds}\/, Annals of mathematical studies,
Vol. 86, Princeton University Press, 1976.

\item{[11]}
D. \spa{McCullough},
{\it Mappings of reducible manifolds}\/,
Geometric and Algebraic Topology; Banach center publications
18 (1986), 61--76.

\item{[12]}
D. \spa{Giulini},
{\it 3-Manifolds for relativists}\/,
Int.~Jour.~Theor.~Phys. 33 (1994), 913--930.

\item{[13]}
H. \spa{Hendriks} and D. \spa{McCullough},
{\it On the diffeomorphism group of a reducible 3-manifold}\/,
Topology~ and its Appl.~ 26 (1987), 25--31.

\item{[14]}
D. \spa{McCullough},
{\it Topological and algebraic automorphisms of 3-manifolds}\/,
in: Groups of Homotopy, Equivalences and Related Topics,
R. Piccinini (ed.), Lecture Notes in Math. 1425, Springer, Berlin,
1990, 102--113.

\item{[15]}
D. \spa{Witt},
{\it Symmetry groups of state vectors in canonical quantum gravity}\/,
Jour.~Math. Phys.~27 (1986), 573--592.

\item{[16]}
J. \spa{Friedman} and D. \spa{Witt},
{\it Homotopy is not isotopy for homeomorphisms of 3-mani\-folds}\/,
Topology~25 (1986), 35--44.

\item{[17]}
A. \spa{Hatcher},
{\it Homeomorphisms of sufficiently large $P^2$-irreducible
3-manifolds}\/,
Topology~15 (1976), 343--347.

\item{[18]}
H. \spa{Hendriks},
{\it Application de la th\'eorie d'obstruction en dimension 3}\/,
Bull.~Soc.~Math. France, suppl\'ement, memoire~53 (1977), 81--195.

\item{[19]}
S.P. \spa{Plotnick},
{\it Equivariant intersection forms, knots in $S^4$, and rotations
in 2-spheres}\/,
Trans.~Am.~Math~Soc.~296 (1986), 543--575.

\item{[20]}
J. \spa{Milnor},
{\it Introduction to algebraic K-theory}\/,
Annals of Mathematics Studies Vol.~72, Princeton University
Press (1971).

\item{[21]}
H.S.M. \spa{Coxeter} and W.O.J. \spa{Moser}\/,
{\it Generators and relations for discrete groups},
(second edition) Springer Verlag, Berlin, G\"ottingen, Heidelberg,
New York (1965).

\item{[22]}
D. \spa{McCullough} and A. \spa{Miller}\/,
{\it Homeomorphisms of 3-manifolds with compressible boundary},
Mem.~Amer.~Math.~Soc.~344 (1986).

}}
\end